\documentclass{IOS-Book-Article}

\usepackage{mathptmx}
\usepackage{graphicx}
\usepackage{enumitem}
\usepackage{float}
\usepackage[colorlinks]{hyperref}

\newcommand{\ourcode}{\sc{Hy-Nbody}}

\def\hb{\hbox to 10.7 cm{}}

\begin{document}

\pagestyle{headings}
\def\thepage{}

\begin{frontmatter}              % The preamble begins here.

%\pretitle{Pretitle}
\title{Direct $N$-body application on low-power and energy-efficient parallel architectures}

\markboth{}{October 2019\hb}
%\subtitle{Subtitle}

\author[A]{\fnms{David} \snm{Goz}%
\thanks{Corresponding Author: David Goz, ORCID: 0000-0001-9808-2283, INAF-Osservatorio Astronomico Di Trieste,
Via G.B. Tiepolo 11, 34131 Trieste, Italy;
E-mail:
david.goz@inaf.it}},
\author[B]{\fnms{Georgios} \snm{Ieronymakis}}
,
\author[B]{\fnms{Vassilis} \snm{Papaefstathiou}}
,
\author[B]{\fnms{Nikolaos} \snm{Dimou}}
,
\author[A]{\fnms{Sara} \snm{Bertocco}}
,
\author[A]{\fnms{Antonio} \snm{Ragagnin}}
,
\author[A]{\fnms{Luca} \snm{Tornatore}}
,
\author[A]{\fnms{Giuliano} \snm{Taffoni}}
and
\author[A]{\fnms{Igor} \snm{Coretti}}

\runningauthor{D. Goz et al.}
\address[A]{INAF-Osservatorio Astronomico di Trieste, Italy}
\address[B]{FORTH-ICS, Heraklion, Crete, Greece}

\begin{abstract}

The aim of this work is to quantitatively evaluate the impact of computation 
on the energy consumption on ARM MPSoC platforms, exploiting CPUs, embedded GPUs and FPGAs.
One of them possibly represents the future of High Performance Computing systems: 
a prototype of an Exascale supercomputer. Performance and energy measurements are made using a state-of-the-art direct $N$-body code from the astrophysical domain. 
We provide a comparison of the time-to-solution and energy delay product metrics, 
for different software configurations.
We have shown that FPGA technologies can be used for application kernel acceleration and are emerging as a promising alternative to ``traditional'' technologies for HPC, which purely focus on peak-performance than on power-efficiency.

\end{abstract}

\end{frontmatter}
\markboth{October 2019\hb}{October 2019\hb}

\section{Introduction and motivation}
\label{section:intro}
Energy efficiency is one of the main problems for exascale computing systems,
since simply re-scaling the current petascale systems would require an unfeasible amount of power consumption. A re-design of the underlying technologies (i.e., processors, interconnect, storage, and accelerators) is needed to reduce energy requirements by about one order of magnitude \cite{Hammer}.
To exploit the upcoming new architectures, software developers are forced to face the challenge of re-designing algorithms.

Commodity single board platforms are an interesting case of heterogeneous systems for performance and energy-efficiency studies (e.g. \cite{Calore_2015,Nikolskiy_2016,Nikolskiy_2018, Morganti_2016}). They are based on low-power System-on-Chip (SoC) architectures with embedded CPUs, GPUs, FPGAs, memory, storage and general purpose I/O ports.
Many companies are delivering single-board computers equipped with different hardware components and utilize Multi-processing System-on-Chip (MPSoC) where the energy efficiency is the main concern.

This work arises in the framework of the ExaNeSt and EuroExa European funded projects aiming at the design and development of a prototype of an exascale HPC facility based on ARM SoC and FPGA technology \cite{exanest1,exanest2}. Our goal is to study the trade-off between time-to-solution and energy-to-solution using real code, a direct $N$-body solver for astrophysical simulations, instead of benchmarking these machines by means of standard suites (e.g. HPL \cite{hpl}, DGEMM, STREAM \cite{stream}). We assess the performance and the associated power-efficiency across different platforms, namely, the MPSoC Firefly-RK3399 produced by Rockchip, and the Zynq-7000 SoC and Zynq UltraScale+ MPSoCs both produced by Xilinx.
We further compare these results with a commodity architecture based on an x86 Intel desktop equipped with a high-end gaming GPU.

To the best of our knowledge, this work provides the first comprehensive evaluation of a real application, coming from the astrophysical domain, on low-cost and low-power boards hosting ARM (64 bit) mobile-class cores, embedded GPUs and FPGAs.

The paper is organized as follows. In Section~\ref{section:code} we describe the code and discuss strategies in order to optimize algorithms on heterogeneous platforms. In Section~\ref{section:platforms} we present the computing platforms used in the test. Section~\ref{section:methodology} is devoted for the discussion of the methodology adopted to make the performance and energy tests. In Section~\ref{section:performances} we present the performance measurements for all platforms along with the power consumption. The last section is dedicated to the conclusions and future work.

\section{$N$-body astrophysical code}
\label{section:code}
In astrophysics, the $N$-body problem is the problem of predicting the individual motions in a group of celestial objects interacting with each other gravitationally. The main drawback related to the direct $N$-body problem relies on the fact that the algorithm requires $O(N^{2})$ computations.
Our application, called {\ourcode}, a modified version of a GPU $N$-body code \cite{nbody:4,nbody:5,nbody:6}, is based on high order Hermite integration schema \cite{hermite} and has been developed in the framework of the ExaNeSt project \cite{exanest1}. {\ourcode} has been designed to fully exploit the compute capabilities of heterogeneous platforms. Three versions of the code are available: one written in Standard C, cache-aware designed for CPUs, one that is implemented and optimized using OpenCL kernels, allowing us to exploit any OpenCL-compliant device (e.g. GPUs), and one that is written also in Standard C using Xilinx Vivado High Level Synthesis (HLS) tool and is implemented for FPGAs.

Code profiling shows that, during a single time step of the simulation, more than 90\% of time is spent on a single kernel with an arithmetic intensity $I \simeq 10^{4}$ [FLOPs/byte] (ratio of FLOPs to the memory traffic), using $32^{3}$  particles. In the following, time and energy measurements on a given device refer to this compute-bound kernel.

\subsection{Floating point arithmetic considerations}
\label{subsection:EX}
The Hermite $6^{th}$ order integration schema requires double precision (DP) floating-point arithmetic in the evaluation of inter-particle distance and acceleration in order to minimize the round-off error, so as to preserve the total energy and the angular momentum of the $N$-body system during the simulation.

Full IEEE-compliant DP floating-point arithmetic is efficient in contemporary CPUs, but it is still extremely resource-eager and performance-poor in other accelerators like gaming or embedded GPUs.
As an alternative, the extended-precision (EX) (or emulated double precision) numeric type \cite{ex} can represent a trade-off in porting {\ourcode} on devices not specifically designed for scientific calculations. An EX-number provides approximately 48 bits of mantissa at single-precision exponent ranges.

\section{Computing platforms}
\label{section:platforms}
\begin{figure}[h]
\centering\includegraphics[width=\linewidth]{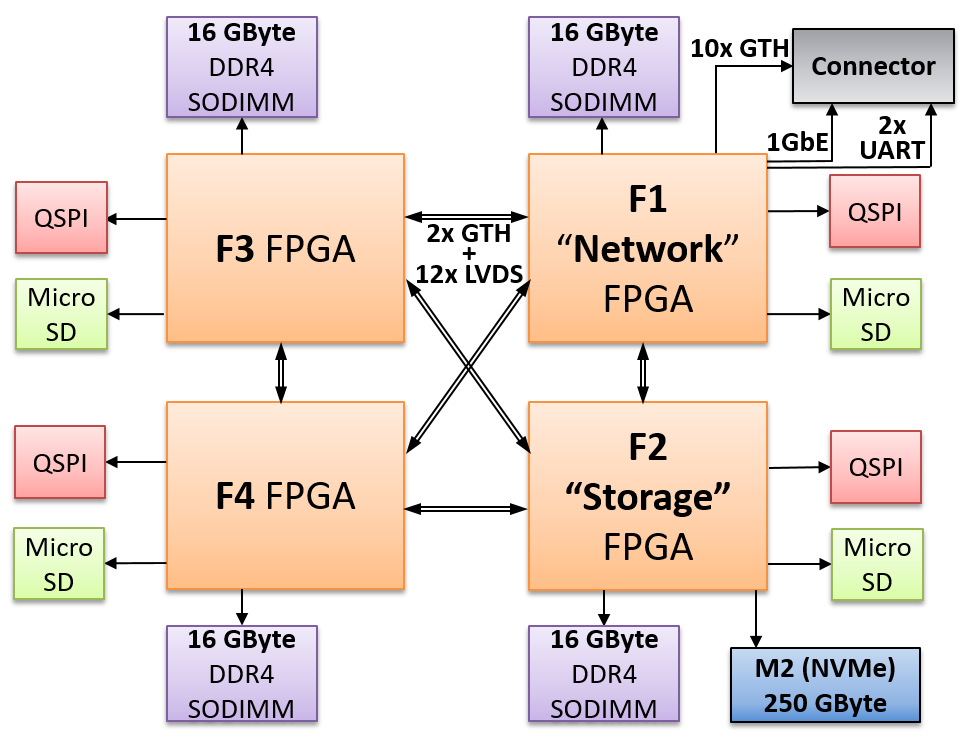}
\caption{The Quad-FPGA daughterboard block diagram and interconnects.}
\label{fig:QFDB}
\end{figure}

In this section we describe the four computing platforms used in our tests.
Table~\ref{table:platforms} lists the devices present in each computing platform, and we highlight in bold the devices used in our tests. The platforms are:

\begin{itemize}
    \item \textbf{Firefly-RK3399 board}: it is equipped with the ARM big.LITTLE architecture, four Cortex-A53 cores with 32KB L1 cache and 512KB L2 cache, and a cluster of two Cortex-A72 high performance cores with 32KB L1 cache and 1MB L2 cache. Each cluster operates at independent frequencies, ranging from 200MHz up to 1.4GHz for the LITTLE, and up to 1.8GHz for the big. The board contains 4GB DDR3 - 1333MHz RAM. The board contains also the OpenCL-compliant Mali-T864 embedded GPU;
    
    \item \textbf{x86 desktop}: it is equipped with four Intel i7-3770 cores running at 3.4 GHz with 32KB L1 cache, 256KB L2 cache and 8192KB L3 cache. The board contains 16GB DDR3 - 1866 MHz RAM and the NVIDIA GeForce-GTX-1080 GPU in the PCI Express (16X) bus;
    
    \item \textbf{ZedBoard}: it is equipped with the Xilinx Zynq 7000 MPSoC, with dual-core ARM Cortex-A9 processors integrated with 28nm Artix-7 based programmable logic (FPGA). The board contains 512 MB DDR3 RAM;
    
    \item \textbf{QFDB}: the Quad-FPGA DaughterBoard (QFDB) is the compute-unit of the prototype developed within the framework of the ExaNeSt project \cite{exanest1, qfdb}.
    The compute board, whose block diagram is shown in Figure~\ref{fig:QFDB}, contains four Xilinx Zynq UltraScale+ MPSoC devices (ZU9EG), each featuring four Cortex-A53 and two Cortex-R5 cores, along with a rich set of hard IPs and Reconfigurable Logic. A 16GB DDR4 RAM is connected to each Zynq device.
    The maximum sustained power of the board is 120 Watts. Targeting a compact design, the dimension of the board is  120-130mm while no component on top or below the printed circuit board (PCB) is taller than 10mm. The PCB stackup consists of 16 layers, with Megtron-6 dielectric. Within the board, multiple high-speed serial links (HSSLs) connect the four Zynq devices, each operating at a line rate of 16.375Gbps. One of the Zynq devices is connected to the outside world through 10 high-speed serial links (HSSLs) using GTH transceivers. On each QFDB, the measurement of the power consumption is accomplished by using a set of TI INA226 current/power sensor, coupled with high-power shunt resistors.
    The INA226 sensor's minimal capture time is 140$\mu s$. However, the Linux INA drivers (and the power-on set-up) set the capture time to 1.1ms by default. The Linux driver also enables averaging from 16 samples and captures both the shunt and the bus voltages. 
    To collect data from the sensors, each board includes 15 I2C power sensors, which allow measuring the power consumption of the primary sub-systems of the board.
   
\end{itemize}

Table~\ref{table:devices} shows the clock, the theoretical peak performance in FP64/FP32 and the achieved performance of the devices using DP/EX arithmetic.
Since FPGAs do not have a fixed architecture, a generic way to calculate
their peak performance does not exist for the following reasons:
\begin{enumerate}[label=(\roman*)]
    \item each type of calculation needs a different amount of resources to be implemented;
    \item a single type of calculation can be implemented in various ways;
    \item the FPGA can operate with various clock frequencies;
    \item usually an accelerator takes a part of the FPGA and not the entire FPGA (even a design of 90\% utilization is very difficult to be placed and routed and it becomes even more difficult in higher clock frequencies).
\end{enumerate}

Thus, in Table~\ref{table:devices}, regarding the FPGAs, we present the theoretical performance of the implemented kernels versus the actual performance obtained including the latency of the memory I/O and the time needed to handle the kernels from the
software application.

\begin{table*}[h]
\caption{The platforms and the associated devices. The devices exploited in the test are highlighted in bold.}
\label{table:platforms}
\begin{center}
%\begin{adjustbox}{width=\textwidth}
\begin{tabular}{|c|c|c|c|}

\hline
\textbf{Platform}  & \textbf{CPU}                 & \textbf{GPU}                     & \textbf{FPGA} \\
\hline
Firefly-RK3399     & \textbf{ARM A53x4 + A72x2}   & \textbf{ARM Mali-T864}           & None \\
\hline
Desktop            & \textbf{Intel i7-3770x4}     & \textbf{NVIDIA GeForce-GTX-1080} & None \\
\hline
ZedBoard           & ARM A9x2                     & None                             & \textbf{Zynq-7000}\\
\hline
QFDB               & 4x(ARM A53x4 + R5x2)             & 4x(ARM Mali-400)                     & 4x(\textbf{Zynq-US+})\\
\hline

\end{tabular}
%\end{adjustbox}
\end{center}
\end{table*}

\begin{table*}[h]
\caption{The clock, the theoretical peak performance and the achieved performance of the devices tested in this work. The actual performance has been obtained using 65536 particles.}
\label{table:devices}
\begin{center}
%\begin{adjustbox}{width=\textwidth}
\begin{tabular}{|c|c|c|c|c|c|c|c|}

\hline
\textbf{Device}  & \textbf{i7x4} & \textbf{A53x4} & \textbf{A72x2} &        \textbf{GTX-1080} & \textbf{T864} &                        \textbf{Z-7000} & \textbf{Z-US+} \\

\hline
Clock (GHz)                            & 3.40   & 1.42         & 1.80      & 1.733      & 0.800 & 0.100 & 0.300 \\
\hline
Peak FP64/FP32 (GFLOPS)    & 108/217   & 11.36/45.44  & 7.2/28.8  & 277.3/8873 & 32/109 &8.5/39.1 &102/351.9
\\
\hline
Actual DP/EX (GFLOPS)   & 8.0/4.1   & 2.6/2.6   & 3.0/2.6   & 55.2/4984   & 1.5/14.4   & 1.2/4.9   & 95.3/333.1\\
\hline

\end{tabular}
%\end{adjustbox}
\end{center}
\end{table*}

\section{Methodology}
\label{section:methodology}
In this section we discuss how power measurements were made. 
On the first three platforms, namely, the Firefly-RK3399, the desktop and the ZedBoard, the electric power draw is measured by means of a power meter (Yokogawa WT310E), while on the QFDB it relies on the on-board sensors. 

After booting up the platform, we measure the watt-hours consumed in idle during a period of three minutes ($\Delta T_{3}$), giving us the $E_{baseline}$ of the system. Then, $E^{device}_{baseline}$ is the electric power drawn by the system running a given code implementation using a particular device (CPU, GPU or FPGA) over $\Delta T_{3}$.

The energy-to-solution of the specific device is:
\begin{equation}
    E^{device}_{impl} = \left(E^{device}_{baseline} - E_{baseline}\right) \cdot \left(T^{device} / \Delta T_{3}\right),
\end{equation}
where $T^{device}$ is the kernel running time (time-to-solution averaged over ten runs). We point out that the benchmark runs have been done taking into account the output to the main memory,  as happens in real production runs.

We also estimate the total energy impact of the application in terms of Energy Delay Product (EDP), as suggested by Cameron \cite{edp}, and defined as:
\begin{equation}
    EDP = E^{device}_{baseline} \cdot \left(T^{device} / \Delta T_{3}\right)^{w},
\end{equation}
where $w$ is a parameter to weight performance versus power (usually $w= 1,2,3$). The EDP is a ``fused'' metric to evaluate the trade-off between time-to-solution and energy-to-solution.

\section{Computational performances and energy consumption}
\label{section:performances}

\begin{figure}[!ht]
\centering\includegraphics[width=\linewidth]{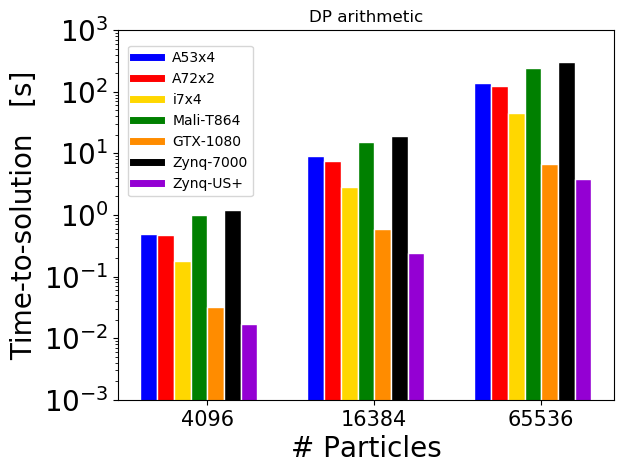}
\caption{Time-to-solution in seconds for different devices as a function of the number of particles for double-precision arithmetic.}
\label{fig:comparison}
\end{figure}

First we investigate the time-to-solution running the code varying the number of particles. In the case of CPUs, we exploit all the available cores by means of OpenMP threads. On the Firefly-RK3399 board equipped with the big.LITTLE ARM architecture, we pinned the processes first to the four cores of the Cortex-A53 and then to the two cores of the Cortex-A72. Kernel execution times on GPUs, both Nvidia-GeForce-GTX-1080 and ARM Mali-T864, have been obtained using the OpenCL's built-in profiling functionality, which allows the host to collect runtime information, while in the case of QFDB the kernel was executed only on a single FPGA out of the four it comprises.

In Figure~\ref{fig:comparison}, we compare the time-to-solution for the devices reported on Table~\ref{table:platforms} for DP arithmetic. From a pure performance point of view, regarding the DP arithmetic, the Zynq UltraScale+ FPGA and the Nvidia GPU are the most powerful devices, while the FireFly MPSoC and the Zynq-7000 FPGA performances are almost two orders of magnitude lower.

\begin{figure}[!ht]
\centering\includegraphics[width=\linewidth]{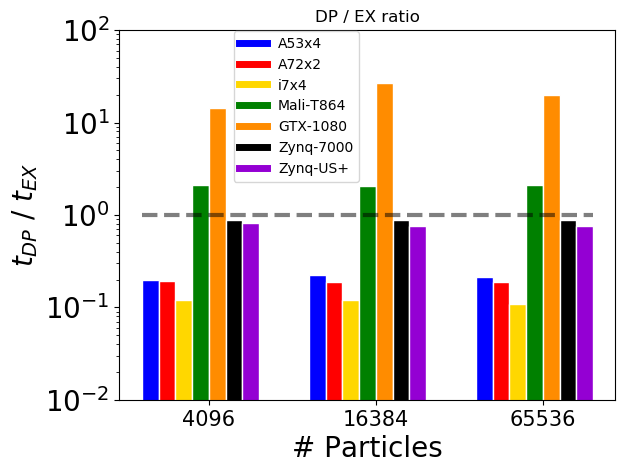}
\caption{The ratio of the time-to-solution between DP arithmetic and EX arithmetic as a function of the number of particles for different devices.}
\label{fig:ratio_DP_EX}
\end{figure}

To better study the effect of the extended-precision arithmetic, in  Figure~\ref{fig:ratio_DP_EX} we show the ratio of time-to-solution between DP and EX arithmetic. The performance improvement is a factor of $\sim 2$ for the Mali-T864 GPU and $\sim 20$ for the GTX-1080, while CPUs suffer a significant performance degradation.
Regarding the QFDB, the EX kernel shows a $32\%$ degradation in performance compared to the DP implementation. Although single precision arithmetic requires less FPGA resources overall, the extra calculations (in particular accumulations) needed to minimize the roundoff and overflow errors in the intermediate results of EX precision introduce an additional overhead which impacts the size of the kernel that can be implemented inside the FPGA. Thus, although the EX kernel was designed to execute on $\sim 25\%$ less particles per cycle, because of these extra calculations it occupies 10\% and 8\% more in terms of DSPs and LUTs accordingly compared to the DP implementation.

\begin{figure}[!ht]
\centering\includegraphics[width=\linewidth]{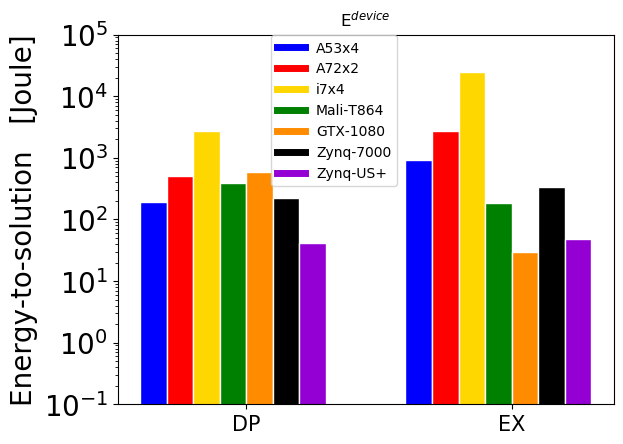}
\caption{The energy-to-solution of the specific device is shown.}
\label{fig:energy_device}
\end{figure}

Figure~\ref{fig:energy_device} shows the total energy-to-solution ($E^{device}$) for all devices and for both DP and EX arithmetic using 65536 particles. For CPUs and GPUs, the instantaneous power consumption is pretty much the same running the DP or EX kernel implementation, however using EX arithmetic on GPUs leads to better energy-efficiency because of the reduced time-to-solution. Our findings show that the Zynq UltraScale+ FPGA is more energy-efficient than the GTX-1080 by a factor of 15 when using DP arithmetic. 

\begin{figure}[!ht]
\centering\includegraphics[width=\linewidth]{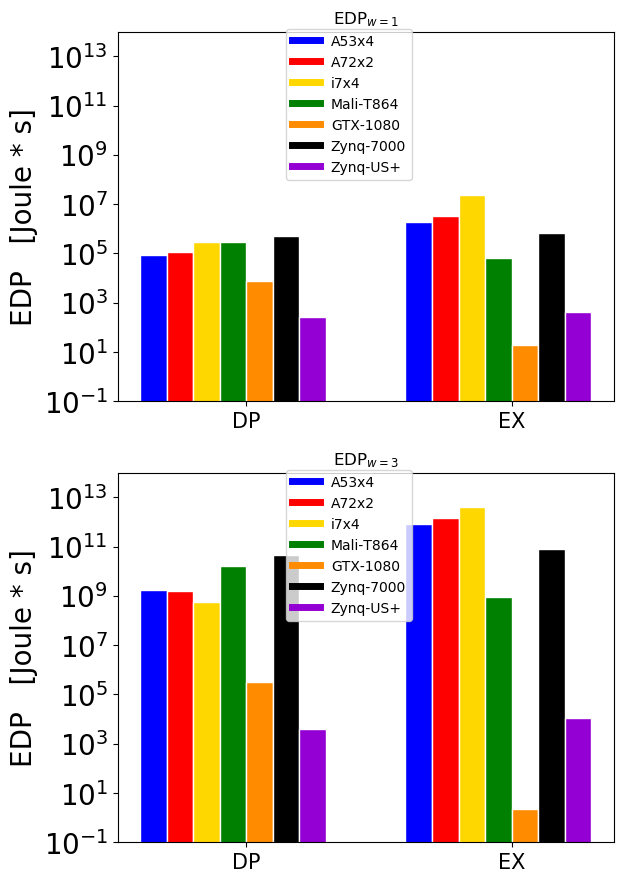}
\caption{EDP for DP arithmetic and for EX arithmetic varying the device. Top panel for $w = 1$, and bottom panel for $w = 3$.}
\label{fig:EDP}
\end{figure}

Moreover, we obtain EDP, for $w= 1,3$, when running the application using 65536 particles and with the methodology discussed in Section~\ref{section:methodology}.
In Figure~\ref{fig:EDP} we plot the EDP comparing the devices (top panel for $w = 1$, and bottom panel for $w = 3$). When performances are highly valued (i.e. $w = 3$), the GTX-1080 is the device with the best trade-off between time-to-solution and energy-to-solution using EX arithmetic, while the Zynq UltraScale+ FPGA is favorable in terms of energy and execution time when DP floating-point arithmetic is used.

\section{Conclusions and future work}
\label{section:conclusions}
The energy footprint of scientific applications will become one of the main concerns in the HPC sector. In this work we employ a real scientific application, coming from Astrophysical domain, 
to explore the impact of software design on time-to-solution and energy-to-solution using low-cost MPSoC-based platforms and FPGA-based technologies that can be potentially used in future HPC systems.

Due to the computational intensive nature of our application, accelerators, like GPU and FPGA, outperform CPU peak performance, as expected. 
In particular, the introduction of the emulated-double precision improves the application performance on SoCs with embedded GPUs, like the ARM Mali, opening the path for successful and cost-effective use of such devices in HPC.

The crucial findings of this work are the achieved performances, both in terms of time-to-solution and energy-to-solution, exploiting the 
Zynq UltraScale+ FPGA on the ExaNeSt QFDB prototype. Kernel development for FPGAs is slightly different than traditional GPU development in that the hardware is created for the specific functions being implemented. Understanding the difference between SIMD parallelism and pipeline parallelism employed on FPGAs, and taking advantage of FPGA features, such as heterogeneous memory support, channels, loop pipelining and unrolling, are key factors to unlock high performance-per-watt solutions.

In conclusion, we have shown that SoC technology is emerging as a promising alternative to ``traditional'' technologies for HPC, which purely focus on peak-performance rather than power-efficiency. The main drawback is that programmers of scientific applications will have to re-engineer their code in order to fully exploit new computing facilities based on heterogeneous hardware.

Our future plan is to assess the energy footprint of the {\ourcode} application on a cluster of MPSoCs, hosting CPUs, GPUs and FPGAs, and to compare it with HPC resources, where multi-node communication becomes an important aspect of the application.

\section{Acknowledgments}
This work was carried out within the EuroEXA and ExaNeSt FET-HPC projects (grant no. 754337 and no. 671553), funded by the European Union's Horizon 2020 research and innovation program.
%\\
%\\
%\\
%\\

\end{document}